# DISCRETE SCALE RELATIVITY

Robert L. Oldershaw

Amherst College

Amherst, MA 01002

USA

rloldershaw@amherst.edu

**ABSTRACT:** The possibility that global discrete dilation invariance is a fundamental symmetry principle of nature is explored. If the discrete self-similarity observed in nature is exact, then the Principle of General Covariance needs to be broadened in order to accommodate this form of discrete conformal invariance, and a further generalization of relativity theory is required.

## I. INTRODUCTION

The Einstein field equations of General Relativity can be written[1] as:

$$\mathbf{R}_{\mu\nu} - \tfrac{1}{2} \mathbf{g}_{\mu\nu} \mathbf{R} = k \mathbf{T}_{\mu\nu} \qquad (1)$$

where $\mathbf{R}_{\mu\nu}$ is the Ricci tensor, $\mathbf{g}_{\mu\nu}$ is the metric tensor, $\mathbf{R}$ is the Ricci scalar, $\mathbf{T}_{\mu\nu}$ is the stress-energy tensor and k is the coupling factor between the geometry of a space-time and its matter content. This equation can be written in an even more compact form:

$$\mathbf{G}_{\mu\nu} = k \mathbf{T}_{\mu\nu} \qquad (2)$$

where $\mathbf{G}_{\mu\nu}$ is called the Einstein tensor. This deceptively simple expression disguises the fact that the equation represents a complicated and coupled set of 10 nonlinear partial differential equations in 4 unknowns. However, the conceptual meaning of the equation has a simple elegance. The geometry of the space-time ($\mathbf{G}_{\mu\nu}$) is determined by the energy and momentum densities/fluxes of the matter ($\mathbf{T}_{\mu\nu}$) and, reciprocally, the motions of the matter are determined by the geometry of the space-time.

In the present paper we explore the implications of a new paradigm for understanding nature, called the Self-Similar Cosmological Paradigm[2-5] (SSCP), and its role in a possible extension of General Relativity. The SSCP draws attention to the self-evident, but often inadequately appreciated, fact that nature is organized in a hierarchical manner, from the smallest observable subatomic particles to the largest cosmological structures. Although the whole hierarchy appears to be quasi-continuous, the SSCP emphasizes that the cosmological hierarchy is *highly stratified*. While the *observable portion* of the entire hierarchy encompasses nearly 80 orders of magnitude in mass, three narrow mass ranges, each extending for only about 5 orders of magnitude, account for ≥ 99% of all mass observed in nature. These dominant mass ranges: roughly $10^{-27}$ g to $10^{-22}$ g, $10^{28}$ g to $10^{33}$ g and $10^{38}$ g to $10^{43}$ g, are referred to as the Atomic, Stellar and Galactic *Scales*, respectively. They constitute the discrete self-similar scaffolding of the observable portion of the quasi-continuous hierarchy. The SSCP proposes that nature's hierarchy extends far beyond our current observational limits on both large and small scales, and is probably unbounded in terms of scale, such that there are no largest or smallest objects (or Scales) in nature.

The SSCP further proposes that the Atomic, Stellar and Galactic Scales, and all other fundamental Scales of nature's infinite hierarchy, are rigorously interrelated by a new symmetry property referred to as *discrete cosmological self-similarity*. For each class of particle, composite system or phenomena on any given Scale, it is proposed that there is a discrete self-similar analogue on all other Scales.[2-5] Mass (M), length (L) and time (T) parameters associated with analogues on neighboring Scales $\Psi$ and $\Psi$-1 are related by the following set of discrete self-similar transformation equations.

$$L_\Psi = \Lambda L_{\Psi-1} \qquad (3)$$

$$T_\Psi = \Lambda T_{\Psi-1} \qquad (4)$$

$$M_\Psi = \Lambda^D M_{\Psi-1} \qquad (5)$$

where $\Lambda$ and D are empirically derived[2,3] dimensionless constants with values of 5.2 x $10^{17}$ and 3.174, respectively. The value of $\Lambda^D$ = 1.70 x $10^{56}$. The symbol $\Psi$ is used as a discrete index for keeping track of specific Scales, and numerically:

$$\Psi = \{\ldots, -2, -1, 0, +1, +2, \ldots\}. \qquad (6)$$

Usually the Atomic, Stellar and Galactic Scales are assigned $\Psi = -1$, $\Psi = 0$ and $\Psi = +1$, respectively. Any dimensional "constant" or parameter appearing in an equation associated with a given Scale must be transformed according to the scaling equations (3), (4) and (5) in order to find the correct value for the counterpart "constant" or parameter on a neighboring Scale.

According to the SSCP, the Newtonian gravitational constant G (= 6.67 x $10^{-8}$ cm$^3$/g sec$^2$) in General Relativity must be replaced by the factor:

$$G_\Psi = [\Lambda^{1-D}]^\Psi G, \qquad (7)$$

where again: $\Lambda = 5.2 \times 10^{17}$ and $D = 3.174$. We may also write $G_\Psi = [\Lambda^{-2.174}]^\Psi G$, since $G \propto L^3/MT^2$ and, according to the discrete self-similar scaling rules of the SSCP, G would therefore scale as $\Lambda^3/\Lambda^D\Lambda^2 = \Lambda^{-2.174}$. It can be seen that for $\Psi = 0$, the expression $[\Lambda^{1-D}]^\Psi G$ reduces to G, in agreement with observations for Stellar Scale gravitational interactions. The possibility that the Atomic Scale gravitational coupling factor is on the order of $10^{38}$ times larger than its counterpart on the Stellar Scale has recently found support in demonstrations that the Atomic Scale value of $G_\Psi$ leads to successful retrodictions of the proton mass and the alpha particle radius.[6]

A revised form of General Relativity that incorporates discrete cosmological self-similarity, which is a form of discrete conformal invariance known as discrete dilation invariance, would be:

$$\mathbf{R}_{\mu\nu} - \tfrac{1}{2}\, \mathbf{g}_{\mu\nu}\, \mathbf{R} = 8\pi/c^4 \,[\Lambda^{1-D}]^{\Psi}\, G\, \mathbf{T}_{\mu\nu} \qquad (8)$$

Instead of a single equation for General Relativity, we now have an infinite set of identical equations, one for each cosmological Scale. Another way to express this is the following equation:

$$\mathbf{R}_{\mu\nu} - \tfrac{1}{2}\, \mathbf{g}_{\mu\nu}\, R = 8\pi/c^4 \, G_{\Psi}\, \mathbf{T}_{\mu\nu}, \qquad (9)$$

where $G_{\Psi}$ is the infinite discrete series of correct SSCP coupling constants for gravitational interactions, one "constant" for each cosmological Scale.

## II. EXTENDING THE PRINCIPLE OF GENERAL COVARIANCE

General Relativity is covariant under conformal transformations (which preserve angles and length ratios, but do not involve absolute length scales), including dilation invariance, when masses and G are suitably scaled. Maxwell's equations of electromagnetism are likewise covariant under conformal transformations, including dilation invariance, when electric charges are suitably scaled. From the point of view of the SSCP, the fact that General Relativity and electromagnetism are both consistent with discrete global dilation invariance is encouraging, since this form of discrete conformal invariance fits very well with the conceptual properties of the SSCP and its discrete self-similar Scale transformation equations.

The Principle of General Covariance states that the laws of physics should be fully independent of arbitrary choices of reference frames and coordinate systems. From this one can infer that the laws of physics are independent of: spatial location, time, spatial orientation, and state of motion (inertial or accelerated). Before the advent of the SSCP, size or scale (note small s) was the one unique thing that was *not* relative, but rather appeared to be absolute. The hydrogen atom in its ground state was thought to have just one fixed set of scale values (m ≈ 1.67 x $10^{-24}$ g, r ≈ 0.53 x $10^{-8}$ cm , t ≈ 1.5 x $10^{-16}$ sec, …). The SSCP, especially in the exact cosmological self-similarity form, proposes a radical change in our thinking about the concept of scale. *Within* a cosmological Scale, one can still invoke absolute scale, but absolute scale no longer applies generally to nature's infinite, discrete hierarchy of self-similar Scales. If each Scale is exactly self-similar to any other Scale, then there are an infinite number of differently sized hydrogen atoms (one for each Scale). Each of these hydrogen atoms has a unique and equally valid set of scale parameters ($m_\Psi$, $r_\Psi$, $t_\Psi$, …).

According to this radically revised natural philosophy, the Principle of General Covariance must now be extended so that the laws of physics are independent of: location, time, orientation, state of motion and *discrete cosmological Scale*. The latter addition reflects the idea that the laws of physics should not depend upon our arbitrary choice of a particular cosmological Scale as our reference system. This is equivalent to saying that the fundamental laws of physics are identical on all cosmological Scales, *for the case of exact cosmological self-similarity*.

## III. DISCRETE SCALE RELATIVITY

The discrete global dilation invariance discussed above is *formally equivalent* to invariance with respect to a discrete global transformation of length, time and mass *units*.

In order to make this clear, we start with the assumption of exact cosmological self-similarity. This means that each cosmological Scale is identical (except for relative scale) in terms of the objects comprising the Scale and their dynamical interactions. Given unlimited observational capabilities, ideal observers on any Scale would describe identical Universes with cosmological self-similarity that is discrete, global and unbounded.

1. <u>Fixed Units Approach</u>: If we want to measure and compare populations of analogue systems from different Scales in terms of one fixed set of units (say the conventional Stellar Scale cm, g and sec that we are familiar with), then the *numerical values* of all length, time and mass parameters, as well as all dimensional "constants", scale according to the discrete self-similar Scale transformation equations of the SSCP. For example, a Stellar Scale proton analogue would have a radius of about $4.2 \times 10^4$ cm and an Atomic Scale proton would have a radius of about $0.8 \times 10^{-13}$ cm, differing by the now familiar scale factor of $\Lambda = 5.2 \times 10^{17}$ for lengths (spatial and temporal).

2. <u>Relative Units Approach</u>: Again we choose the cm, g and sec as our length, mass and time units, but now we acknowledge that there is a *different* set of cm, g and sec for each Scale of the infinite discrete cosmological hierarchy. Because the discrete self-similarity is exact, each [$cm_\Psi$, $g_\Psi$, $sec_\Psi$] set is *identical*, *equally fundamental* and *equally valid*. One can easily see that reductionism is an endless "fool's errand" in this paradigm. In the relative units approach, the discrete self-similar Scale transformation equations tell us the *relative* "sizes" of these *units*. For example, the Stellar Scale proton and the Atomic

Scale proton now both have the *same* radius of about $4.2 \times 10^4$ $cm_\Psi$, but the Atomic Scale centimeters ($cm_{\Psi=-1}$) are $5.2 \times 10^{17}$ times smaller than the Stellar Scale centimeters ($cm_{\Psi=0}$).

Within the context of this discrete scaling of relative units, *all fundamental physical laws and their constants are identical for each Scale, in both form and numerical values*. This relativity of Scale can be designated as Discrete Scale Relativity, or more simply as Scale Relativity, so long as we remember that:

$$\text{Scale} \equiv \text{cosmological Scale} \neq \text{scale}.$$

Scale Relativity is a further generalization of General Relativity, and the hierarchical subsuming of relativity theories can be symbolically expressed:

**{[(S**pecial Relativity**)** General Relativity**]** Scale Relativity**}**.

A more comprehensive and detailed discussion of this discrete self-similar cosmological paradigm, including its development, predictions, retrodictions, publications, theoretical implications, etc., can be found at the author's website:

http://www.amherst.edu/~rloldershaw .